# Mobile Augmented Reality to Discover New Environments

Nehla GHOUAIEL    Jean-Marc CIEUTAT    Jean-Pierre JESSEL

ESTIA-IRIT    ESTIA-IRIT    IRIT

n.ghouaiel@estia.fr    j.cieutat@estia.fr    jean-pierre.jessel@irit.fr

## Abstract

Although man has become sedentary over time, his wish to travel the world remains as strong as ever. The aim of this paper is to show how techniques based on imagery and Augmented Reality (AR) can prove to be of great help when discovering a new urban environment and observing the evolution of the natural environment. The study's support is naturally the Smartphone which in just a few years has become our most familiar device, which we take with us practically everywhere we go in our daily lives.

## Introduction

The term augmented reality was first used in 1992 by Tom Caudell and David Mizell to name the overlaying of computerized information on the real world. Subsequently, the expression was used by Paul Milgram & Fumio Kishino in their seminal paper "Taxonomy of Mixed Reality Visual Displays" [13]. In this paper, they describe a continuum between the real world and the virtual world (nicknamed mixed reality) where augmented reality evolves close to the real world whereas augmented virtuality evolves close to the virtual world (figure 1).

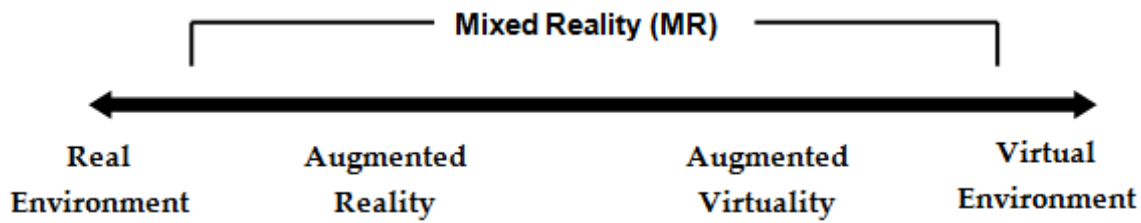

Figure 1. Continuum between reality and virtual reality

In 1997 Ronald Azuma developed a complementary definition which he completed in 2001 [14] and which, along with Milgram & Kishino's approach, gave two commonly admitted definitions of augmented reality. According to Azuma, an augmented reality system is one which complements the real world with (computer generated) virtual objects so they seem to coexist in the same space as the real world, which in both cases leads him to define the features of an augmented reality system according to the following three properties:
1. "Combining real and virtual". In the 3D real world 3D entities must also be integrated.
2. "Real time interactivity". This namely excludes films even if the previous condition is respected.
3. "3D repositioning". This enables virtual entities to be made to visually coincide with reality.
Displaying augmentations can be done with direct or indirect vision (thus inducing an additional mental load). In the case of direct vision, the display uses metaphors such as mirrors; smartphones open like windows onto the environment, vision through glasses or windows, etc.

Pervasive computing is a technology that enables sensing, computing, advanced electronics and wireless communication to be embedded in everyday objects [15]. As computing devices become progressively smaller and more powerful, the technology is moving beyond the personal computer to other devices.

Pervasive computing is a larger field than augmented reality. Pervasive computing can use Augmented Reality technology as a possible way of augmenting the users' environment. The computing and interaction hardware required for augmented reality becomes smaller and more available in everyday life, which is the case of Smartphone. By consequence, mobile Augmented Reality and pervasive computing can converge, creating systems that are ubiquitously available and allow interaction in the style of augmented reality.

In this paper we show how augmented reality can be a pervasive tool enabling people finding out new environments, indoor as well as outdoor. In the first part of this paper, we present our proper definition of augmented reality. The second part of this paper summarizes our work on mobile outdoor augmented reality. It presents a sensor-based graphic application for urban navigation and an image-based technique showing environmental changes through the ages. The third part details marker and markerless object recognition techniques for augmented reality. Further in this section, we depict the architecture of our mobile system for object recognition.

## 1. Our definition of Augmented Reality

All the definitions proposed leave little room for multimodality. However, augmented reality has today exceeded the stage of repositioning virtual indices in a video flow and now also proposes sound and even tactile augmentations. In [6] we proposed a more general definition of augmented reality as being the combination of physical spaces with digital spaces in semantically linked contexts. We can say that augmented reality is the combination of physical spaces with digital spaces in semantically linked contexts for which the objects of associations lie in the real world. On the contrary, we can define augmented virtuality as being the combination of physical spaces with digital spaces in semantically linked contexts, but where the task's objects lie in the world of computing, states that the systems considered aim to make interaction more realistic.

## 2. Mobile Outdoor Augmented Reality
### 2.1. Urban Environment

Tourists visiting an urban environment for the first time, may face a number of problems. They may, for example, not initially have a precise destination [2]. On the other hand, in any urban environment there are Points of Interest (POIs), which visitors may easily miss if these are less well known or difficult to locate. This type of POI may be described as hidden. D.

McGookin [2] shows how visitors can pass by statues without actually seeing them. In this case, the first issue facing tourists confronted with unfamiliar urban environments is: What is worth visiting in the city? We believe that the most appropriate answer to this question in such situations, should at least contain all the POIs (the most interesting places to visit in urban environments in this case) with highest priority ranking. Priority ranking POIs are those situated close to the visitor's position as well as those considered to be the city's symbols (this is the case of the Eiffel Tower in Paris). To distinguish common land navigation point by point (in which the destination is determined) from navigation in which the destination is not known in advance, we have chosen to call the latter multipoint navigation.

One of the aims of augmented reality is to enhance perception or the visibility of the physical world. The Smartphone's screen acts as a window onto the real world whose video flow can be augmented. We use the geo-referenced data of objects to inform users about their location as shown in figure 2, for example, where the location information of different POIs located close by can be seen. Our system calculates the user's position based on GPS data. It then filters the database so as to only display POIs close to the user. Filtering calculates the distance between the user and the referenced objects using the Haversine formula [16].

$$a = \sin^2(\Delta\varphi/2) + \cos(\varphi 1).\cos(\varphi 2).\sin^2(\Delta\lambda/2) \quad (1)$$
$$c = 2.\operatorname{atan2}(\sqrt{a}, \sqrt{(1-a)}) \quad (2)$$
$$d = R.c \quad (3)$$

Where $\varphi$ is latitude, $\lambda$ is longitude, R is earth's radius (mean radius = 6,371km). With regard to the display, annotations are added to the real scene, which are visible on the smartphone's screen as illustrated in figure 2. For this purpose, we use the "Vision See Through (VST)" technique [3], widely used in augmented reality applications. Just like the documented reality functionality relating to augmented reality, our video flow can be enriched with information identifying what can be seen with the camera.

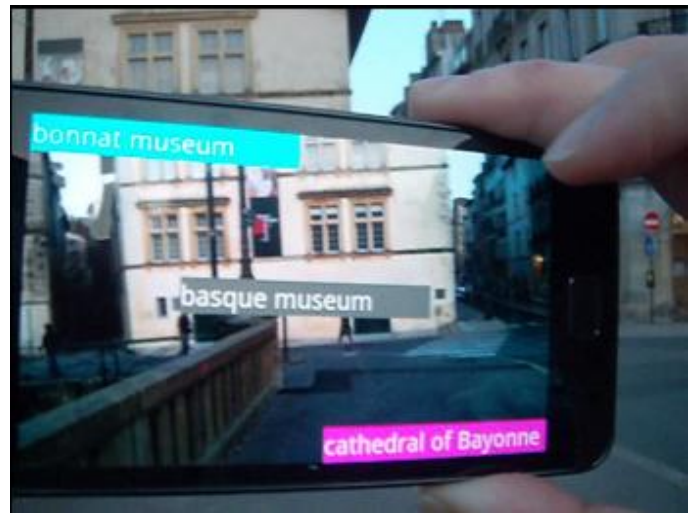

Figure 2. Visual interface of our augmented reality browser

The layout of annotations informs users about the spatial location of POIs with regard to their geographical position. For example, the annotation in the top left means that the POI in question is in front of the user on the left.

## 2.2. Natural Environment
### 2.2.1. Related work

Environmental changes are a subject of interest for researchers, managers and the public at large. Changes in the environment affect people's lives and therefore are a subject of great concern. J. Danado et al [4] present a mobile system enabling the quantity of water and the pollution levels in artificial lakes and natural rivers to be visualised. The system proposed is based on a client server architecture and consists of two modules: an augmented reality module and a geo-referencing module. The geographical information system (GIS) was used in addition to augmented reality [5], to modellise natural landscapes and shows their evolution over time. The case study presented in [5] illustrates the propagation of weeds.

### 2.2.2. Visualization of landscape transformations

Urbanisation over recent decades and the growing interest in new technologies, have led people to become estranged from nature and to no longer know their own environment. Furthermore, over time natural environments suffer transformations which continually modify their forms. This is the case of the cliffs in Hendaye and Etretat in figure 6. Thus, it is very interesting to enable visitors to a natural environment to visualise these changes. In this context, visitors may, for example, hold their devices (PDA, Smartphone, etc.) and see what the landscape looked like in the past, from their point of view. In technological terms, several image-based techniques are capable of modellising the natural environment's transformations. An illustration of coastal erosion concerns a smart ''morphing'' algorithm. The contours between the sea, the coastline and the sky are initially taken from source image and destination image (figure 3); then, ''morphing'' animation intermediate images are then calculated by interpolation based on contour coordinates by applying the following equation:

$$I_\alpha = I_s * \alpha + I_d * (1 - \alpha) \quad (4)$$

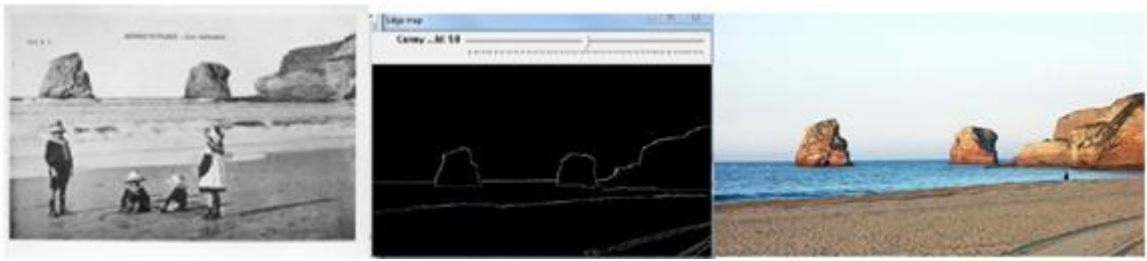

(a) Source Image      (a) edge Image      (c) last Image

**Figure 3. Morphing Computing**

Thus, the illusion of a rendering, which faithfully retranscribes the retreat of the coastline over time due to the effect of the sea, is therefore perfect.

3. **Indoor Augmented Reality**
    3.1. **QR Codes**

QR Code is acronym for Quadratic Residue Code [12]. QR code is a two dimensional symbol. It was invented in 1994 by Denso, one of major Toyota group companies, and approved as an ISO international standard (ISO/IEC18004) in June 2000. QR Code is a matrix type symbol that has a cell structure arranged in a square. It consists of the functionality patterns for making reading easy and the data area where the data is stored. QR Code contains the following elements: finder patterns, alignment patterns, timing patterns, and a quiet zone.

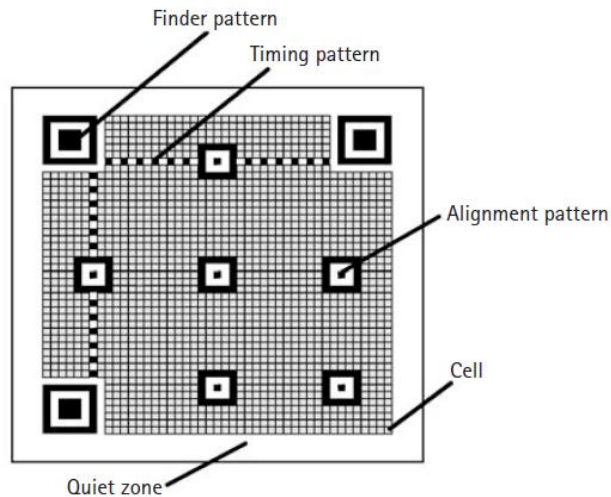

**Figure 4.Structure of QR Code**

QR Code data is stored into the data area. The grey part in Figure 11 represents the data area. The data area is encoded into the binary numbers of '0' and '1' based on the encoding rule. The binary numbers of '0' and '1' are converted into black and white cells and then arranged. The data area has Reed-Solomon codes incorporated for the stored data and the error correction functionality.

3.2. **Artworks Augmentation**

Figure 5 illustrates the application we implement to augment Artworks. When the user flash the QR code, a descriptive text appears, overlaid on the camera view.

To highlight the advantage of our AR application in favoring culture learning, we carry out a user experience. 16 unpaid subjects take part in the experimental study; they are 8 males and 8 females. Subjects are aged between 22 and 40 years old. The task consists of contemplating 20 pictures of artworks. We add QR codes to only 10 pictures. We stuck descriptive etiquettes under the rest of pictures. At the end of experience, we ask each subject two questions, one is related to a random artwork associated with a QR code and the other is related to an artwork with descriptive etiquette. Results shows that 65% of subjects succeed in answering a question related to a QR code but only 27% succeed in answering the question related to etiquette. The obtained results allow us to conclude that AR may help museum visitors to retain more information about artworks and other exhibited objects. Therefore, it is worth to encourage the use of AR application in discovering cultural and historic heritage inside museum.

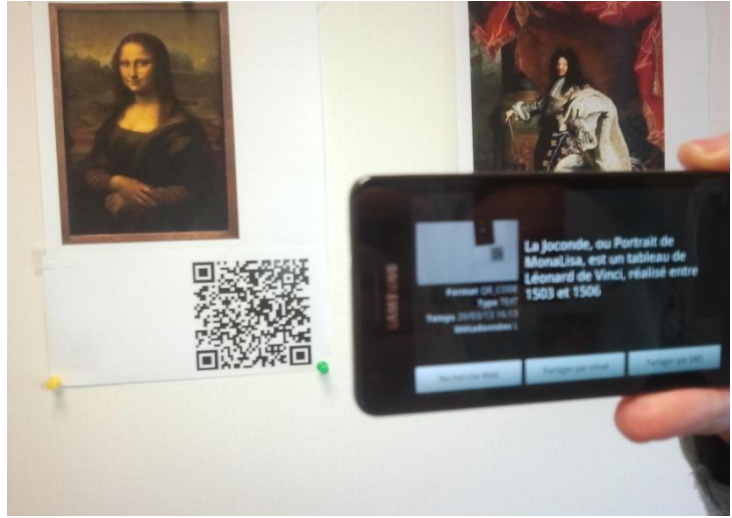

**Figure 5. Artwork Augmentation**

### 3.3. Features points

The use of QR code results in a kind of visual pollution. Hence, in this section we describe alternative solution to QR code which is features points. The task of finding point correspondences between two images is a part of object recognition. The most known detectors are the Harris corner detector [7], proposed back in 1988. Harris corner detector uses the eigenvalues of the second moment matrix. However, Harris corners suffer from scale variance.

SIFT detector introduced by Lowe [8] is a scale-invariant detector. This detector captures a substantial amount of information about the spatial intensity patterns, while at the same time being robust to small deformations or localization errors. The descriptor detailed in [8], called SIFT, computes a histogram of local oriented gradients around the interest point and stores the bins in a 128 - dimensional vector (8 orientation bins for each of 4 * 4 location bins). SURF point features detector, is derived from SIFT. It was first presented in 2006 as a novel scale and rotation-invariant detector and descriptor. It shares with SIFT the same concept of local features descriptors based on the neighbourhood of the interest point. Nevertheless, SURF differs in how the interest points are selected and described. SURF [9] detector is based on the Hessian matrix because of its good performance in computation time and accuracy. It relies on the determinant of Hessian matrix for selecting the location and the scale of a feature point. Given a point x = (x, y) in an image I, the Hessian matrix H(x, σ) in x at scale σ is defined as follows:

$$H(x, \sigma) = \begin{bmatrix} L_{xx}(x, \sigma) & L_{xy}(x, \sigma) \\ L_{xy}(x, \sigma) & L_{yy}(x, \sigma) \end{bmatrix} (5)$$

Where Lxx(x, σ) is the convolution of the Gaussian second order derivative $\frac{\partial^2}{\partial x^2} g(\partial)$ with the image I in point x, and similarly for Lxy(x, σ) and Lyy(x, σ). The extraction of SURF descriptor is performed in two steps. The first step consists of finding the orientation to a circular region around the interest point. Then, a square region aligned to the selected orientation is constructed, and therefore the SURF descriptor is extracted from it. Thanks to the use of integral images, SURF detector is faster than others point features detectors. An integral image can be rapidly computed from an input image and used to speed up the computation of the SURF descriptors for that image. The value of the integral image I(x) in a point (x,y) is the sum of all the pixel values of the input image I between the point and the origin.

$$I_\Sigma(x) = \sum_{i=0}^{i \leq x} \sum_{j=0}^{j \leq y} I(i,j) \quad (6)$$

The integral image enables fast computation of the intensities over any upright rectangular area of the image. This process is independent of the size of the image or of the area.

### 3.4. Object Recognition with SURF

Once visual features have been extracted from an image, they are matched against a set of features extracted from other images. All the feature descriptors covered in the previous section (see 2.1) contain a vector of real numbers. The simplest way to compare two features is to compute the Euclidean distance (or the squared Euclidean distance) between descriptors, in n-dimensional space. This computation is obviously slower if the dimension is higher, so descriptors with smaller vector (like the 64-dimensional SURF) are preferable over larger ones (like the 128-dimensional SIFT). The distance between two vectors p and q is evaluated using an Euclidean metric:

$$dist(p,q) = \sqrt{\sum_{i=1}^{64}(p_i - q_i)^2} \quad (6)$$

Our feature matching algorithm is based on Fast Approximate Nearest Neighbour Search Algorithm [10][11]. It is an n-n search algorithm that uses randomised KD-Trees to give an approximate nearest neighbour index. The approach adapted by kd-tree algorithm consists of hierarchically decomposing space into a relatively small number of cells. By consequence, each node contains few input objects. This enables us to access any input object by position. We sweep down the hierarchy until we reach the cell containing the object. Kd-trees algorithm is constructed by partitioning point sets recursively along across different dimensions. Plane through one dimension defines a node, also called child, in the tree. Nodes are partitioned into equal halves, using planes through a different dimension. Partitioning process stops after log n levels, with each point in its own leaf cell. Median point is used for partition.

However, to avoid comparing the entire n feature extracted with the m features in the database (with a complexity of nm), the sign of the Laplacian can be used for fast matching. As the sign of the Laplacian discerns dark blobs on light background from light blobs on dark background, there is no need to compare two features with a different sign as they will not be related to the same feature.

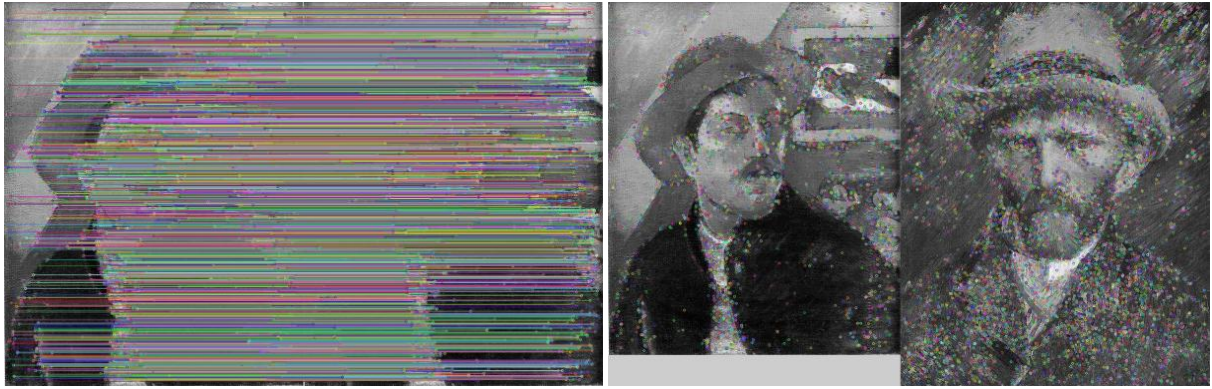

Figure 6. Portraits matching

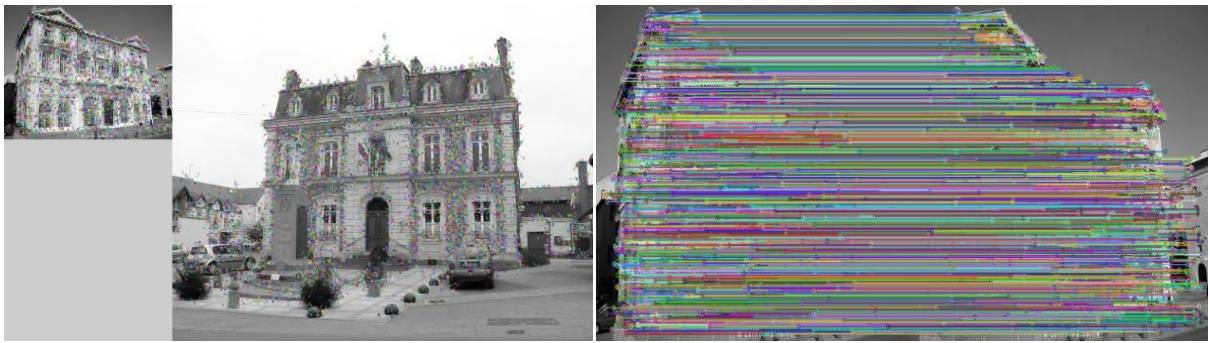

Figure 7. Building matching

### 3.5. Database creation

We created a database of features in which we record, for each object, all SURF descriptors. When matching a feature extracted from a test image with the database, a comparison between features is performed. Our approach considers only the sign of the Laplacian and the descriptor vector, in evaluating the dissimilarity between two features; the others elements of the descriptors are ignored. The figure below shows the database schema.

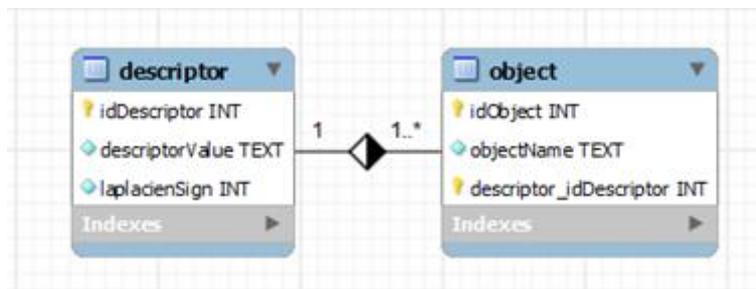

Figure 8. database schema

### 3.6. Mobile Recognition System

In many of augmented reality applications, we need to recognize objects, in order to apply correspondent augmentations on them. In the context of pervasive augmented reality, recognition process should be omnipresent. For this purpose, we choice client/server architecture, illustrated by the schema below. This architecture exploits inherent characteristics of cloud computing to provide omnipresent recognition service.

Database of features is stored in a distant server. Mobile scans the environment and sends image of captured object to a remote server. The server extracts SURF features from the image and then seeks for correspondences using the algorithm detailed in section 3.4.

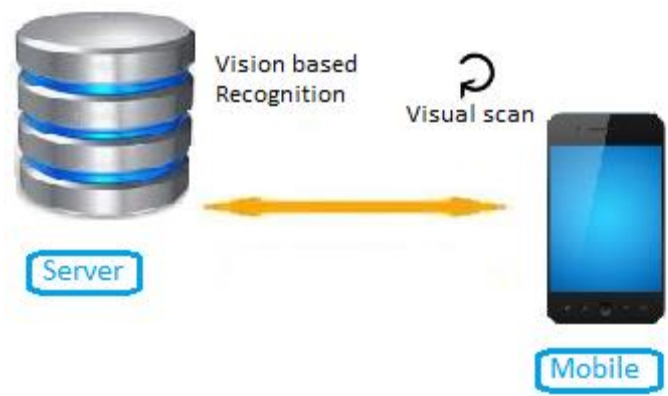

**Figure 9. Schema of pervasive recognition system**

## Conclusion

Augmented reality is the combination of physical spaces with digital spaces in semantically linked contexts for which the objects of associations lie in the real world. Pervasive computing is a technology that enables sensing, computing, advanced electronics and wireless communication to be embedded in everyday objects [15]. Mobile Augmented Reality and pervasive computing can converge, creating systems that are omnipresent and allow interaction in the style of augmented reality. The concept of pervasive augmented reality is illustrated by indoor and outdoor applications. In the case of this paper, we show the use of augmented reality for helping visitors discover new environments.

Object recognition is a primordial process in augmented reality, notably in the case where augmentations are expected to be applied everywhere in real world. The object must be known in order to apply exact augmentation to it. Hence, at the end of this paper, we depict our method for object recognition, used in our mobile augmented reality systems.

As mobile devices become progressively more powerful, we envisage transform recognition system from client server architecture to self-contained server architecture. In fact, self-contained server architecture may speed up system response, given that latency of network communication no longer exists.